\documentclass{SCGE}
\usepackage{multicol}
\usepackage{graphicx}

\newcommand{\mgii}{Mg\,{\footnotesize II}} %{\ion{Mg}{2}}
\newcommand{\ha}{H\ensuremath{\alpha}}
\newcommand{\hb}{H\ensuremath{\beta}}
\newcommand{\civ}{C\,{\footnotesize IV}} %{\ion{C}{4}}
\newcommand{\feii}{Fe\,{\footnotesize II}}
\newcommand{\oiii}{O\,{\footnotesize III}}
\newcommand{\mbh}{\ensuremath{M_\mathrm{BH}}}

\begin{document}

\begin{picture}(0,0){\rm
\put(0,-39){\makebox[160truemm][l]{\bf {\sanhao\raisebox{2pt}{.}}
Research Paper  {\sanhao\raisebox{1.5pt}{.}}}}}
\end{picture}

\def\bm{\boldsymbol}

\def\dl{\displaystyle}
\def\du{\end{document}}

\Year{2012} %
\Month{000}
\Vol{000} %
\No{000} %
\BeginPage{000} %
\EndPage{000} %
\AuthorMark{{\rm WANG Jianguo,} et al.}
\DOI{000} % The author doesn't need fill in it.

% \title[short text for running head]{full title}{comments for title}
\title{On the Systematic Bias in the Estimation of Black Hole Masses in Active Galactic Nuclei}%标题

\author[1,2,3*]{WANG Jianguo}{}%换手动
\author[4,5]{Dong Xiaobo}{}

\address[{\rm1}]{National Astronomical Observatories/Yunnan
Observatory, Chinese Academy of Sciences, Kunming 650011, China;}
\address[{\rm2}]{Key Laboratory for the Structure and Evolution of Celestial
Objects, Chinese Academy of Sciences, Kunming 650011, China;}
\address[{\rm3}]{Graduate University of Chinese Academy of Sciences, Beijing
100049, China}
\address[{\rm4}]{Key Laboratory for Research in Galaxies and Cosmology, University
of Science and Technology of China (USTC), Hefei 230026, China}
\address[{\rm5}]{Center for Astrophysics, USTC, Hefei 230026, China}

\maketitle \vspace{-3.5mm}{\footnotesize\begin{center} Received Apr. 26, 2012; accepted 000%收稿日期
\end{center}}\vspace*{-5mm}

%     Abstract is required.
\begin{center}
\rule{16.5cm}{0.4pt}
\parbox{16.5cm}
{\begin{abstract}
In this report, we find the \mbh\ estimated from the formalism of
Wang et al. (2009)[1] are more consistent with those from the
\mbh-$\sigma_*$ relation than those from previous single-epoch mass
estimators, using a large sample of AGNs. Furthermore, we examine the
differences between the line widths of \hb\ and \mgii\ in detail by
comparing their line profiles. The flux around the line core and that
in the wing of both \hb\ and \mgii\ show an opposite variation tendency,
which indicates the BLR is multi-componential. The contribution of the
wing makes the FWHM deviate from $\sigma_{line}$, and thus bias the \mbh\
estimated from previous single-epoch mass estimators. Thus the
correction on the formalism suggested by Wang et al. (2009)[1]
is crucial to \mbh\ estimation.
\end{abstract}}
\end{center}%\vspace*{-0.6cm}
\begin{center}
\parbox{16.5cm}{\bf\jiuhao Quasars, Galactic nuclei, Masses, Statistical
and correlative studies of properties %关键词
}
\end{center}

\begin{center}
\parbox{16.5cm}{\PACS{\hspace*{-2mm}\rm 98.54.Aj, 98.62.Js, 98.62.Ck, 98.62.Ve}%分类号
\rule{16.5cm}{0.4pt}}\end{center}

%--------------------------------citation---------------------------------------------

%--------------------------------citation---------------------------------------------

%%%%%%%%%%%%%%%%%%%%%%%%%%%%%%%%%%%%%%%%%%%%%%%%%%%%%%%%%%%%
\wuhao\vspace*{1.5mm}
\begin{multicols}{2}
%%%%%%%%%%%%%%%%%%%%%%%%%%%%%%%%%%%%%%%%%%%%%%%%%%%%%%%%%%%%
%% Text of article.
%%%%%%%%%%%%%%%%%%%%%%%%%%%%%%%%%%%%%%%%%%%%%%%%%%%%%%%%%%%%
%    Section headings
\renewcommand{\baselinestretch}{1.08} \baselineskip 12.2pt\parindent=10.8pt

\no %正文

\section{Introduction}
Accretion onto super-massive black holes (SMBHs) is generally
considered as the energy engine of active galactic nuclei (AGNs).
The determination of the mass of SMBH (\mbh ) is crucial to the
understanding of most physical processes associated
with SMBH and the cosmological evolution of black holes.
The \mbh\ of type I AGNs are usually measured using the virial
theorem, \mbh=fR$_{BLR}$V$^2$/G, if the size of broad line
region (R$_{BLR}$) and the virial velocity (V) of clouds in
the BLR are known, where f is a factor of order unity depending
on the geometry and kinematics of the BLR. R$_{BLR}$
can be estimated using the reverberation mapping (RM) method[2],
which monitors the variability of continuum and emission lines.
V can be estimated from the widths of emission lines. Conversely,
\mbh\ can also be estimated using the tight correlation between \mbh\
and the stellar velocity dispersion of the galactic bulge (\mbh-$\sigma_*$
relation)[3,4,5]. However, both of these methods cannot be
used for large samples of AGNs,
\noindent\rule{2.5cm}{0.4pt}\\[0.1mm]{\qihao *Corresponding author (email:
wangjg@ynao.ac.cn)\\
}%手动E-mail地址
\no
because the RM method is time-consuming and the
measurements of $\sigma_*$ are limited by the spectral and saptial
resolution of telescopes.

For large samples of AGNs,  \mbh\ can be estimated
by combining R$_{BLR}$, which is estimated using
the important relationship between R$_{BLR}$ and the
monochromatic continuum luminosity (R-L relation)[6,7,8],
and the FWHM of emission lines. The single-epoch mass estimators
have been studied for various broad lines, such as
\hb\ [1,9], \ha\ [10], \mgii~$\lambda$2800 [1,11,12] and
\civ~$\lambda$1549[13]. If both \hb\ and \mgii\ FWHMs are good
tracers of the virial velocity and can be used to estimated
the \mbh, they should give the same \mbh\ values.
Some researchers found they are consistent with each other [11,14,15,16],
while others came to an opposite conclusion [1,17,18].
Wang et al. (2009) found that \mgii\ FWHM is systematically smaller
than \hb\ FWHM, and that the relationships between \hb\ and \mgii\ FWHM
and $\sigma_{line}$, which is the best virial velocity
tracer measured on the variable part of the spectrum[19],
deviate from the 1:1 relationship. The dependance of \mbh\
on FWHM should be \mbh$\propto$FWHM$^{\gamma}$,
where $\gamma$ is smaller than 2 for both \hb\ and \mgii.
If this is the case, most previous single-epoch mass estimators
(\mbh$\propto$FWHM$^2$) would introduce systematic biases
in \mbh\ estimations[1,20,21,22,23] and result in many artificial
conclusions, as discussed by Rafiee and Hall (2011)[21] and
Croom (2011)[22]. Thus, further testing the validity of the formalism
of Wang et al. (2009)[1] is critical for eliminating such biases
in \mbh\ estimations and many other related relationships in AGNs.
The \mbh\ estimated from the \hb\ and \mgii\ formalisms of
Wang et al. (2009) are more consistent with those from RM
measurements than those from previous single-epoch mass estimators
and are consistent with each other for a large sample culled from
Sloan Digital Sky Survey (SDSS) Data Release 5 (DR5). However,
one remaining issue is whether the new \mbh\ estimates are
consistent with those derived from the \mbh-$\sigma_*$ relation,
which should be tested using a large sample.

Moreover, the reasons for the systematic deviations
between \hb\ and \mgii\ FWHM and $\sigma_{line}$ are
unclear now. The profile of an emission line is determined by
the structure and kinematics of the BLR, which are complex.
It is possible that the broad lines in most AGNs are
generated in multi-regions, including the gravitationally-bound
BLR, outflows[24] and the surface of accretion disk(Wang et al.
2005[25]; Wu et al. 2008[26]). Different measurements
of line width, such as $\sigma_{line}$ and FWHM, would
represent different information about the structure and/or
kinematics of the BLR. Special attention must be noted
when using in the estimation of \mbh. The study of the
structure and kinematics of the BLR would be helpful to
understand why FWHM deviates from $\sigma_{line}$ and important
for the \mbh\ estimation of AGNs. In this report, we examine
whether there are systematic biases between the \mbh\ estimated
from the single-epoch mass estimators and those from the
\mbh-$\sigma_*$ relation. We also compare the profiles of
\hb\ and \mgii\ in order to understand their differences
and why their FWHM deviates from $\sigma_{line}$.

\section{The Bias in the \mbh\ Estimates}
We first verify the consistency between the \mbh\ estimated
from the single-epoch mass estimators and those from the
\mbh-$\sigma_*$ relation (G{\"u}ltekin et al. 2009)[27].
We select 8470 AGNs with $z<0.35$ from SDSS DR4.
The spectrum is corrected for the Galactic extinction
using the extinction map of Schlegel et al. (1998)[28]
and the reddening curve of Fitzpatrick (1999)[29].
The fitting method is described in Dong et al. (2008)[30]
and described below. In the wavelength range
4030-7500\AA, we fit simultaneously the featureless
continuum and the \feii\ multiplets and other emission
lines. Each of the [\oiii]~$\lambda\lambda$ 4959,5007 doublets
is modeled with two Gaussians, one for the line core
and the other for the possible blue wing. The narrow
components of \ha\ and \hb\ are fitted with similar
profile to the line core of [\oiii]~$\lambda$5007 and
the broad components of them are fitted with 1-4 gaussians.
\mbh\ can be estimated using the width of the line
core of [\oiii]~$\lambda$5007 as substitute for
$\sigma_*$[31]. Because \hb\ is weak for many objects,
we estimate \hb\ FWHM from \ha\ FWHM[10] and then
estimate the \mbh\ using the single-epoch mass
estimators. We find that the differences between the \mbh\
estimated from the single-epoch mass estimators and
those from the \mbh-$\sigma_*$ relation are
correlated with \ha\ FWHM (Figure 1), if the formalisms
from Greene and Ho (2005; hereafter GH05)[10] or Vestergaard and Peterson
(2006; hereafter VP06)[9] are used. The correlation would decrease
largely, if the formalism of Wang et al. (2009)[1] is
adopted. The relationship between the \mbh\ differences and FWHM(\ha) is
somewhat linear in log-log space and can be expressed as
$\log \frac{\mbh}{M_{\rm sig}} = \rm k\log \frac{\rm FWHM(\ha)}{\rm km s^{-1}}+$b.
The (k,b) for the formalisms of GH05, VP06 and Wang et al. (2009)
given by the regression method of Kelly (2007)[32] are
(2.05$\pm$0.04, -7.51$\pm$0.13), (1.93$\pm$0.03, -6.73$\pm$0.09) and
(0.86$\pm$0.03, -2.99$\pm$0.11), respectively. All the intrinsic scatters
of these relations are around 0.02 dex. This indicates that
the \mbh\ estimated from the formalism of Wang et al. (2009)[1] are less
biased than those from previous single-epoch mass estimators
(\mbh$\propto$FWHM$^2$). We attempt to estimate the \mbh\ using
the \mbh-$\sigma_*$ relation from other authors (Xiao et al. 2011[33])
and find the \mbh\ estimated from the formalism of Wang et al.
(2009)[1] are still less biased than those from previous
single-epoch mass estimators.

\vspace*{3mm}
\begin{center}
\centerline{\psfig{figure=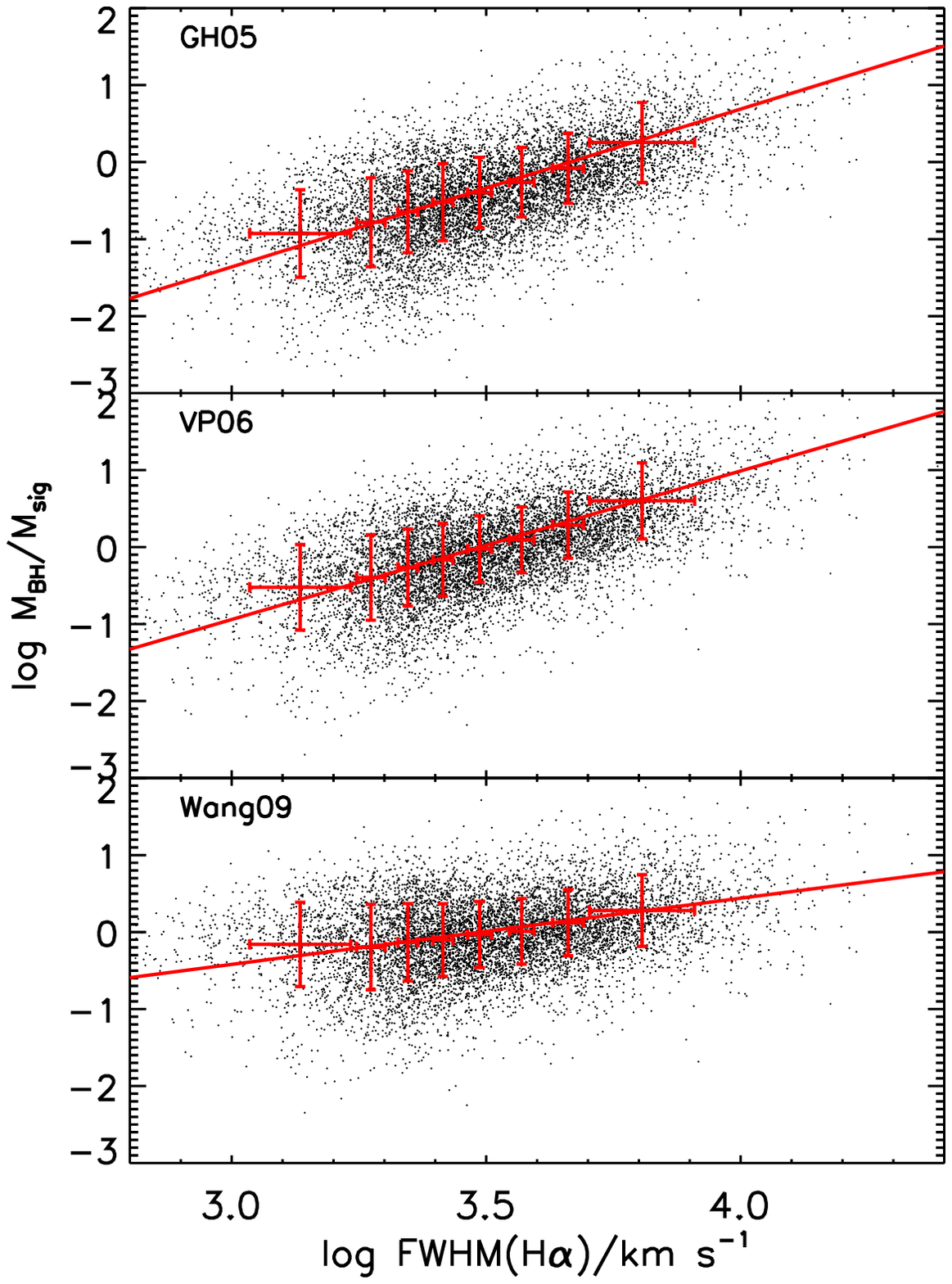,height=8cm,width=7cm}}\end{center}
%\centerline{\psfig{figure=mbhdevfwhm.eps}\end{center}
\vspace*{-5mm} {\footnotesize {\bf Figure 1}\quad
Correlations between FWHM(\ha) and the differences of \mbh\
estimate from single-epoch mass estimators and those
from \mbh-$\sigma_*$ relation[27] for the sample from SDSS DR4.
The crosses are the median values and standard deviations of
the \mbh\ differences and FWHM in each bin of FWHM.
The solid lines show the best-fit relations.
} \vspace*{3mm}

\section{Profiles of \hb\ and \mgii\ }
The comparison above shows that the method of Wang et al.
(2009)[1] is capable of correcting the systematic biases
in the \mbh\ estimations over a large redshift interval.
This indicates indirectly that \hb\ and \mgii\ FWHMs are
deviating from $\sigma_{line}$ systematically. The
systematic deviation may be caused by the complex structure
and kinematics of the BLR. We compare the profiles of
\hb\ and \mgii\ using the sample from Wang et al. (2009)[1],
which was selected from SDSS DR5. The
sample includes 495 AGNs with high signal-to-noise
ratio (S/N~$>$~20) in both the \hb\ (4600-5100 \AA) and the
\mgii\ (2700-2900 \AA) regions, which makes it suitable for
the comparison. The spectrum is corrected for the Galactic
extinction using the extinction map of Schlegel et al. (1998)[28]
and the reddening curve of Fitzpatrick (1999)[29]. The redshifts
of these quasars are from Hewett and Wild (2010)[34],
which were derived by cross-correlating observed spectra with
a carefully constructed template. The dependence of emission line
shift on luminosity and redshift are corrected and the systematic
errors of redshifts are reduced to the level of 30 km/s, which are
important to our investigation. We perform
the continuum and emission-line fitting using an Interactive
Data Language (IDL) code based on MPFIT [35], which performs
$\chi^2$-minimization by the Levenberg-Marquardt technique.

\vspace*{3mm}
\begin{center}
\centerline{\psfig{figure=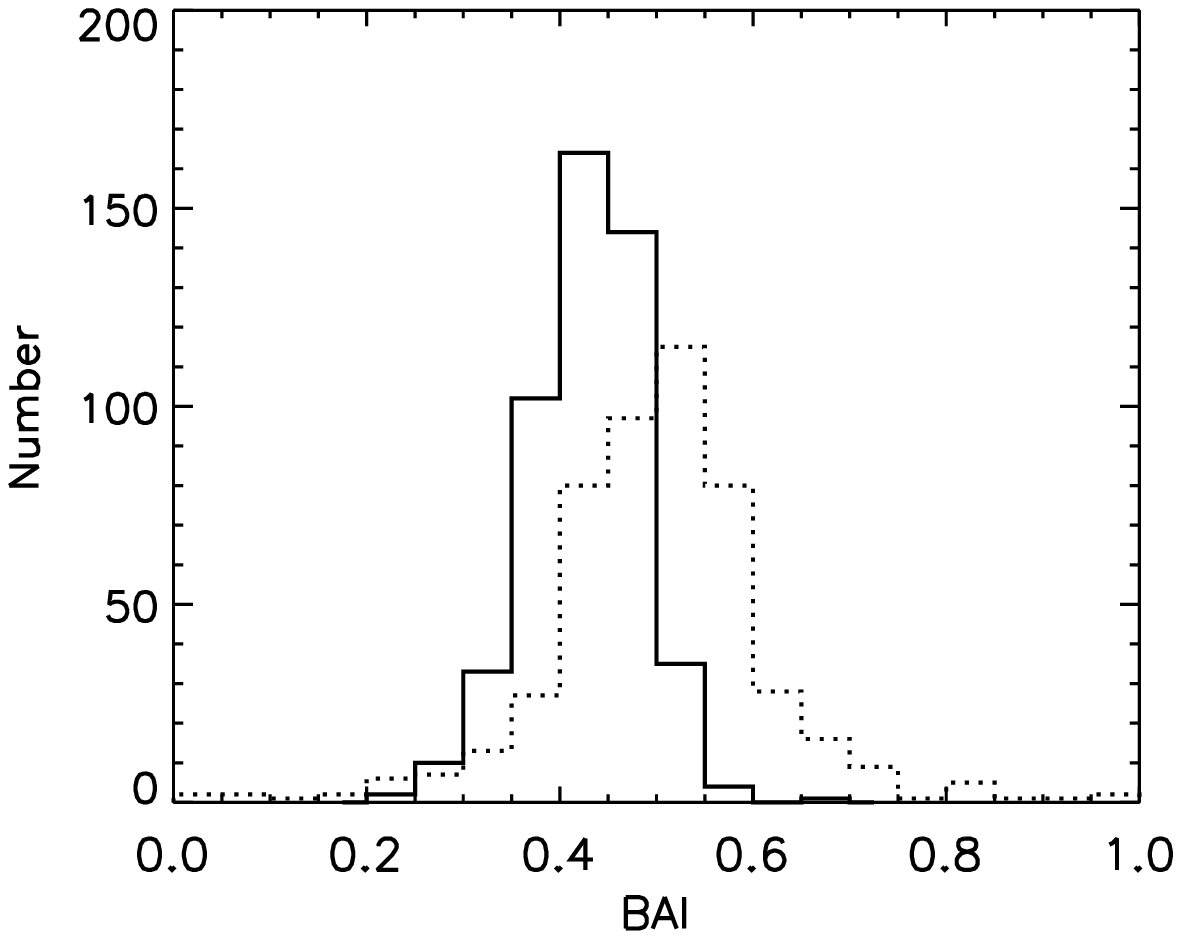,height=6cm,width=7cm}}\end{center}
%\centerline{\psfig{figure=baidis.eps}}\end{center}
\vspace*{-5mm} {\footnotesize {\bf Figure 2}\quad
BAI distributions of \hb\ (solid line) and \mgii\
(dotted line).
} \vspace*{3mm}

The spectrum is fitted in two wavelength range: \hb\ range
(4200-5600\AA ) and \mgii\ range (2200-3500 \AA ). For the
\hb\ range, the fitting method is similar to that described
above. For the \mgii\ range, the method is described in Wang
et al. (2009)[1]. The featureless continuum and \feii\ multiplets
were modeled simultaneously. The broad component of
each of the \mgii$\lambda\lambda$ 2796,2803 doublets is
modeled with a Gauss-Hermite series profile and the narrow
component of each of the doublets is modeled by a Gaussian.

Usually, the shift and asymmetry of lines are studied
separately, while they may be caused by the same process[24].
The blueshift and asymmetry index (BAI), which is defined as
the flux ratio of the blue part to the total profile, measures
their combined effects[24]. For \hb\ and \mgii, the blue part is
the part at wavelength short than 4862.68 and 2800.26 \AA,
respectively. The distributions of BAI are showed in Figure 2.
The median value of the BAI distribution of \mgii\ is around
0.5, while that of \hb\ is smaller than 0.5. Because both \hb\ and
\mgii\ show no evidence of shift (see Figure 3), the BAI is
primarily caused by the line asymmetry. This indicates that \mgii\
profile is quite symmetrical in that there are more flux in the
red part of \hb\ than that in the blue part, which are consistent
with the conclusion if the shift and asymmetry of lines are
measured separately.

\vspace*{3mm}
\begin{center}
\centerline{\psfig{figure=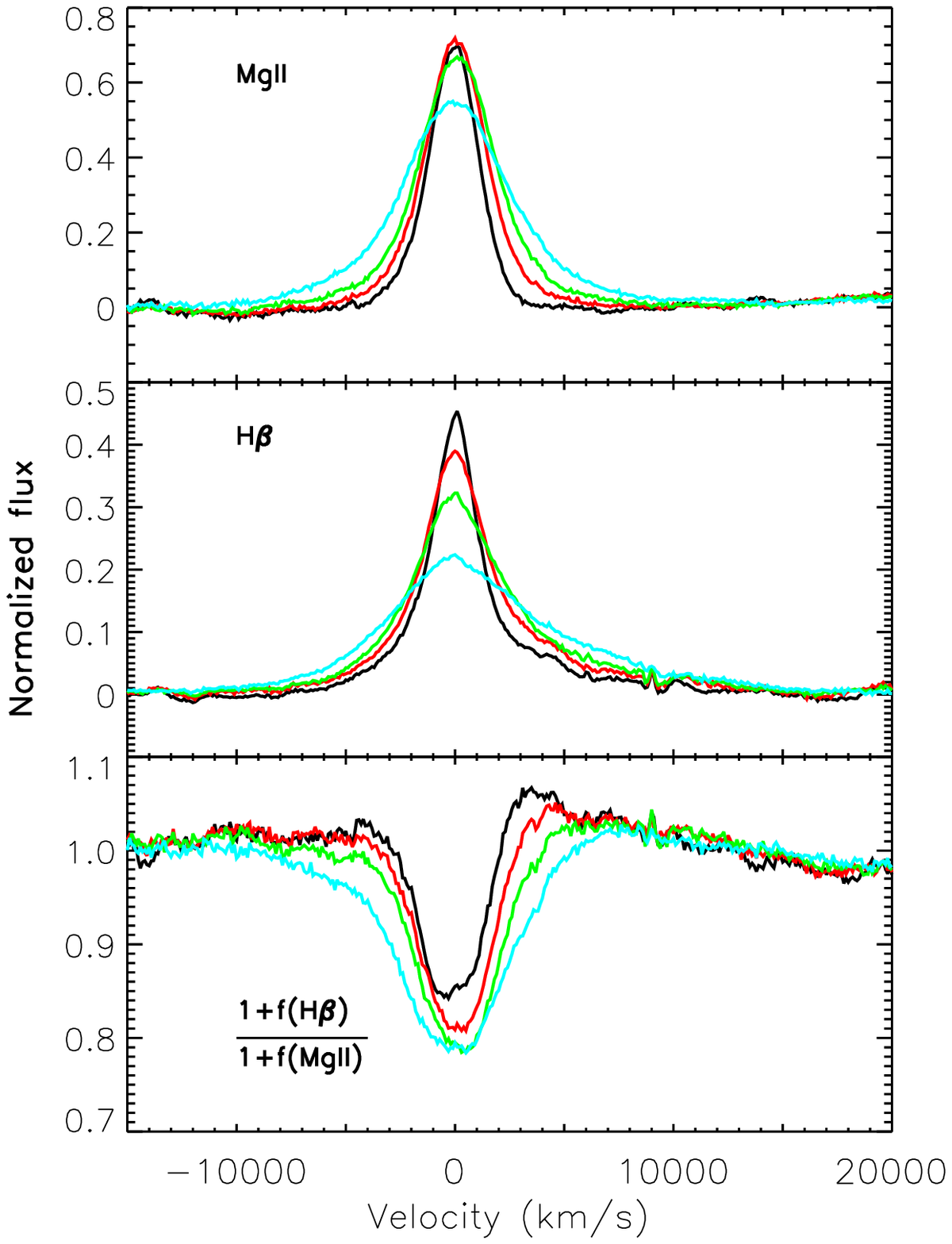,height=10cm,width=7cm}}\end{center}
%\centerline{\psfig{figure=hbmgprocomp.eps}}\end{center}
\vspace*{-5mm} {\footnotesize {\bf Figure 3}\quad
Composite profiles of \hb\ and \mgii\ of four sub-sample divided
by their \hb\ FWHM, as well as their difference in velocity space.
First two panels are \hb\ and \mgii\ profiles. All these flux are
normalized at the emission-line-free window $3030-3090$~\AA\ and the
continuum and \feii\ multiplets were subtracted. The last panel shows
the line ratios of \hb\ and \mgii.
Black: FWHM(\hb) $<$ 3000~km/s; Red: 3000~km/s~$<$~FWHM(\hb)~$<$~4000~km/s;
Green: 4000~km/s~$<$~FWHM(\hb)~$<$~5500~km/s; Cyan: FWHM(\hb)~$>$~5500~km/s.
} \vspace*{3mm}

A direct comparison between the profiles of \hb\ and \mgii\ is showed
in Figure 3. The spectra are normalized at the emission-line-free
window $3030-3090$~\AA\ and the featureless continuum and \feii\
multiplets are subtracted. The sample is divided into four
sub-samples according to \hb\ FWHM. The composite \hb\ and
\mgii\ profiles of each sub-sample, as well as their line ratio,
are showed in Figure 3. The peaks of both \hb\ and \mgii\ do not
show evident shift. The flux in the wings increases with the
increase of FWHM, while the flux around the line core decreases with
the increase of FWHM. The change of \hb\ is more rapid than
that of \mgii. This indicates that \hb\ and \mgii\ are not cospatial
in BLR and the BLR in AGNs is multi-componential. At least
two emitting regions are needed: an intermediate line region (ILR)
producing the line core and a very broad line region (VBLR) producing
the line wings[36]. The emission in ILR makes larger contribution
to the \mgii\ lines, while the emission in the VBLR makes larger
contribution to the Balmer lines[36].

\section{Discussion}
Different structures of the BLR have been proposed to explain the
profiles of emission lines.  These models includes: a rotating
accretion disk, binary black holes, bipolar outflow and
anisotropically illuminated spherical BLR (see Eracleous
and Halpern 2003 and reference therein)[37], as well as the
gravitationally-bound BLR+outflow model of Wang et al. (2011)[24].
The gravitationally-bound BLR+outflow model has succeeded
in explaining the profiles of the high ionization \civ\ line[24],
but is not suitable to explain the profile of low ionization \hb\
line. This is because \hb\ shows a systematically small BAI ($<$0.5) opposed
to the expectation of the model (BAI$>$0.5). Eracleous and Halpern
(2003) found the accretion disk emission could explain the
double-peak profile and other spectroscopic properties of
AGNs presenting the double-peaked Balmer lines, while other
structures are unsatisfactory[37].
They attempted to explain the profiles of \hb\ and \mgii\ using
the accretion disk emission, but the model predicts lower \mgii\ flux
than the observed flux. One of the possible reasons is that
the BLR is two-componential. The contribution of the ILR to \mgii\
is critical but is not included in their model.

As showed in Figure 3, the contribution of the VBLR to \hb\ and
\mgii\ flux becomes more important with the increase of \hb\
FWHM. However, the contribution of the VBLR to $\sigma_{line}$
may be small, because clouds in the VBLR might be optically thin
to the ionization continuum[38]. The emission from the VBLR does not
vary with the variability of continuum and contributes little
to the variable part of the spectrum. This may be the reason
of the systematic deviations between \hb\ and \mgii\ FWHM and
$\sigma_{line}$. Morevoer, the fraction of the contribution
of the VBLR to \mgii\ is much smaller than that to \hb, which makes
the \mgii\ FWHM systematically smaller than \hb\ FWHM.
The contribution of the VBLR makes FWHM deviate from $\sigma_{line}$
systematically and bias the \mbh\ estimation from previous
single-epoch mass estimators ($\mbh\propto$FWHM$^2$). When
estimating the \mbh\ using FWHM as the tracer of the virial velocity,
it is crucial to correct the biases by using the fitted index of the
\mbh$\propto$FWHM$^{\gamma}$ relation, rather than the assumed $\gamma=2$,
as suggested by Wang et al. (2009).

\Acknowledgements{\bahao We thank the anonymous referees for their
helpful suggestions that improved the paper.
We acknowledge useful comments and
suggestions from Weimin Yuan and Chang You.
This work was supported by Chinese NSF (grant Nos. 11073019,
10973034, 11033007, 11133006 and 11103071) and the National Basic Research
Program of (973 Program) 2009CB824800.}


\normalsize \vskip0.3in\parskip=0mm \baselineskip 18pt
\renewcommand{\baselinestretch}{1.1}\footnotesize\parindent=4mm\bahao

\REF{1\ } Wang J G, Dong X B, Wang T G, et al. Estimating Black Hole Masses in Active Galactic Nuclei Using the \mgii\ $\lambda$2800 Emission Line.
Astrophys J, 2009, 707: 1334--1346

\REF{2\ } Blandford R D, McKee C F.  Reverberation mapping of
 the emission line regions of Seyfert galaxies and quasars. Astrophys J,
 1982, 255: 419--439

\REF{3\ } Gebhardt K, Kormendy J, Ho L C, et al.
 Black Hole Mass Estimates from Reverberation Mapping and from Spatially Resolved Kinematics. Astrophys J, 2000, 543: L5--L8

\REF{4\ } Ferrarese L, Pogge R W, Peterson B M, et al.
 Supermassive Black Holes in Active Galactic Nuclei. I. The Consistency of Black
 Hole Masses in Quiescent and Active Galaxies. Astrophys J, 2001, 555: L79--L82

\REF{5\ } Onken C A, Peterson B M, Dietrich M, et al.
 Black Hole Masses in Three Seyfert Galaxies.
 Astrophys J, 2003, 585: 121--127

\REF{6\ } Kaspi S, Smith P S, Netzer H, et al.
Reverberation Measurements for 17 Quasars and the Size-Mass-Luminosity Relations in Active Galactic Nuclei. Astrophys J, 2000, 533: 631--649

\REF{7\ } Bentz M C, Peterson B M, Pogge R W, et al.
The Radius-Luminosity Relationship for Active Galactic Nuclei: The
Effect of Host-Galaxy Starlight on Luminosity Measurements.
Astrophys J, 2006, 644: 133--142

\REF{8\ } Bentz M C, Peterson B M, Netzer H, et al.
The Radius-Luminosity Relationship for Active Galactic Nuclei: The Effect of Host-Galaxy
Starlight on Luminosity Measurements. II. The Full Sample of Reverberation-Mapped AGNs.
Astrophys J, 2009, 697: 160--181

\REF{9\ } Vestergaard M, Peterson B M. Determining Central Black Hole Masses in Distant
Active Galaxies and Quasars. II. Improved Optical and UV Scaling Relationships.
Astrophys J, 2006, 641: 689--709

\REF{10\ } Greene J E, Ho L C. Estimating Black Hole Masses in Active Galaxies Using
the \ha\ Emission Line. Astrophys J, 2005, 630: 122--129

\REF{11\ } McLure R J, Jarvis M J. Measuring the black hole masses of high-redshift
quasars. Mon Not Roy Astron Soc, 2002, 337: 109--116

\REF{12\ } Kong M Z, Wu X B, Wang R, et al. Estimating Black Hole Masses of AGNs using
Ultraviolet Emission Line Properties. China J. Astron Astrophys, 2006, 6: 396--410

\REF{13\ } Vestergaard M. Determining Central Black Hole Masses in Distant Active Galaxies.
Astrophys J, 2002, 571: 733--752

\REF{14\ } Salviander1 S, Shields G A, Gebhardt K, et al.
The Black Hole Mass-Galaxy Bulge Relationship for QSOs in the Sloan Digital Sky Survey
Data Release 3. Astrophys J, 2007, 662: 131--144

\REF{15\ } Hu C, Wang J M, Ho L C, et al.  A Systematic Analysis of \feii\ Emission in
Quasars: Evidence for Inflow to the Central Black Hole. Astrophys J, 2008, 687: 78--96

\REF{16\ } Shen Y, Richards G T, Strauss M A, et al. A Catalog of Quasar Properties from
Sloan Digital Sky Survey Data Release 7. Astrophys J Suppl Ser, 2011, 194: 45

\REF{17\ } Corbett E A, Croom S M, Boyle B J, et al. Emission linewidths and QSO black hole
mass estimates from the 2dF QSO Redshift Survey. Mon Not Roy Astron Soc, 2003, 343: 705--718

\REF{18\ } Dietrich M, Hamann F. Implications of Quasar Black Hole Masses at High Redshifts.
Astrophys J, 2004, 611: 761--769

\REF{19\ } Peterson B M, Ferrarese L, Gilbert K M, et al. Central Masses
and Broad-Line Region Sizes of Active Galactic Nuclei. II. A Homogeneous Analysis
of a Large Reverberation-Mapping Database. Astrophys J, 2004, 613: 682--699

\REF{20\ } Rafiee A, Hall P B. Supermassive Black Hole Mass Estimates Using Sloan
Digital Sky Survey Quasar Spectra at 0.7~$<$~z~$<$~2. Astrophys J Suppl Ser,
2011, 194: 42

\REF{21\ } Rafiee A, Hall P B. Biases in the quasar mass-luminosity plane.
Mon Not Roy Astron Soc, 2011, 415: 2932--2941

\REF{22\ } Croom S M. Do Quasar Broad-line Velocity Widths Add Any Information to
Virial Black Hole Mass Estimates? Astrophys J, 2011, 736: 161

\REF{23\ } Peterson B M. Masses of Black Holes in Active Galactic Nuclei: Implications
for NLS1s. arXiv:1109.4181

\REF{24\ } Wang H Y, Wang T G, Zhou H Y, et al. Coexistence of Gravitationally-bound
and Radiation-driven \civ\ Emission Line Regions in Active Galactic Nuclei. Astrophys J, 2011, 738: 85

\REF{25\ } Wang T G, Dong X B, Zhang X G, et al. Two Extreme Double-peaked Line
Emitters in the Sloan Digital Sky Survey. Astrophys J, 2005, 625: L35--L38

\REF{26\ } Wu S M, Wang T G, Dong X B. Broad reprocessed Balmer emission from warped
accretion discs. Mon Not Roy Astron Soc, 2008, 389: 213--222

\REF{27\ } G{\"u}ltekin K, Richstone D O, Gebhardt K, et al. The M-$\sigma$ and
M-L Relations in Galactic Bulges, and Determinations of Their Intrinsic Scatter.
Astrophys J, 2009, 698: 198--221

\REF{28\ } Schlegel D J, Finkbeiner D P, Davis M. Maps of Dust Infrared Emission for
Use in Estimation of Reddening and Cosmic Microwave Background Radiation Foregrounds.
Astrophys J, 1998, 500: 525

\REF{29\ } Fitzpatrick E L. Correcting for the Effects of Interstellar Extinction.
Publ Astron Soc Pac, 1999, 111: 63--75

\REF{30\ }Dong X B, Wang T G, Wang J G, et al. Broad-line Balmer decrements in blue
active galactic nuclei. Mon Not Roy Astron Soc, 2008, 383: 581--592

\REF{31\ } Komossa S, Xu D. Narrow-Line Seyfert 1 Galaxies and the \mbh-$\sigma$ Relation.
Astrophys J, 2007, 667: L33--L36

\REF{32\ } Kelly B C. Some Aspects of Measurement Error in Linear Regression of Astronomical Data.
Astrophys J, 2007, 665: 1489--1506

\REF{33\ } Xiao T, Barth A J, Greene J E, et al. Exploring the Low-mass End of the
\mbh-$\sigma_*$ Relation with Active Galaxies. Astrophys J, 2011, 739: 28

\REF{34\ }Hewett P C, Wild V. Improved redshifts for SDSS quasar spectra.
Mon Not Roy Astron Soc, 2010, 405: 2302--2316

\REF{35\ }Markwardt C B. Non-linear Least-squares Fitting in IDL with MPFIT.
In: David A, Bohlender D D, Patrick D, eds. Proceedings of the conference held 2-5 November 2008
at Hotel Loews Le Concorde, Qu\'{e}bec City, QC, Canada, 2009, 411: 251--254

\REF{36\ } Popovi{\'c} L {\v C}, Mediavilla E, Bon E, et al. Contribution of the disk emission
to the broad emission lines in AGNs: Two-component model. Astron \& Astrophys, 2004, 423: 909--918

\REF{37\ } Eracleous M, Halpern J P.  Completion of a Survey and Detailed Study of Double-peaked
Emission Lines in Radio-loud Active Galactic Nuclei. Astrophys J, 2003, 599: 886--908

\REF{38\ } Ferland G J, Korista K T, Peterson B M. Optically thin thermal emission
as the origin of the big bump in the spectra of active galactic nuclei.
Astrophys J, 1990, 363: L21--L25

\end{multicols}

\end{document}